\documentclass[letterpaper]{article} % DO NOT CHANGE THIS
\usepackage{aaai24}
\usepackage{times}  % DO NOT CHANGE THIS
\usepackage{helvet}  % DO NOT CHANGE THIS
\usepackage{courier}  % DO NOT CHANGE THIS
\usepackage[hyphens]{url}  % DO NOT CHANGE THIS
\usepackage{graphicx} % DO NOT CHANGE THIS

\usepackage{tabularx}
\usepackage{amsmath}

\usepackage{mathrsfs}
\usepackage{amsfonts}
\usepackage{dutchcal}
\usepackage{multirow}
\usepackage{subfigure}

\usepackage{amsmath,amsfonts}
\usepackage{array}
\usepackage{booktabs}
 
\usepackage[skip=1pt,labelfont=bf]{caption}
 \usepackage{calligra}
\usepackage{color}
\usepackage{colortbl}
\usepackage{courier}
\usepackage{csvsimple}
\usepackage{enumitem}
\usepackage{fancybox}
\usepackage{fontenc}
\usepackage{graphicx}

\usepackage{listings}
\usepackage{longtable}
\usepackage{lscape}
\usepackage{makecell}
\usepackage{marvosym}
\usepackage{moreverb}
\usepackage{multicol}
\usepackage{multirow}
\usepackage{makecell}
\usepackage{pifont}
\usepackage{rotating}

\usepackage[most]{tcolorbox}
\usepackage{threeparttable}
\usepackage{tikz}
\usepackage{soul}
\usepackage{url}
\usepackage{soul}
\usepackage{wasysym}
\usepackage{xspace}
\usepackage{amsthm}

\newcolumntype{L}[1]{>{\raggedright\let\newline\\\arraybackslash\hspace{0pt}}m{#1}}
\newcolumntype{C}[1]{>{\centering\let\newline\\\arraybackslash\hspace{0pt}}m{#1}}
\newcolumntype{R}[1]{>{\raggedleft\let\newline\\\arraybackslash\hspace{0pt}}m{#1}}

\definecolor{codegreen}{rgb}{0,0.6,0}
\definecolor{codered}{rgb}{1,0,0}
\definecolor{codegray}{rgb}{0.9,0.9,0.9}
\definecolor{codepurple}{rgb}{0.58,0,0.82}
\definecolor{backcolour}{rgb}{0.95,0.95,0.92}
\definecolor{lightgray}{gray}{0.9}

\lstdefinestyle{mystyle}{
    commentstyle=\color{codegreen},
    keywordstyle=\color{magenta},
    numberstyle=\small\color{black},
    stringstyle=\color{codepurple},
    basicstyle=\scriptsize\ttfamily,
    breakatwhitespace=false,
    breaklines=true,
    captionpos=b,
    keepspaces=true,
    showspaces=false,
    showstringspaces=false,
    showtabs=false,
    tabsize=2
}

\lstdefinelanguage{diff}{
  morecomment=[f][\color{blue}]{@@},     % group identifier
  morecomment=[f][\color{red}]-,         % deleted lines
  morecomment=[f][\color{codegreen}]+,       % added lines
  morecomment=[f][\color{red}]{---}, % Diff header lines (must appear after +,-)
  morecomment=[f][\color{codegreen}]{+++},
}

\setlist{noitemsep} %to leave space around whole list

\definecolor{darkpastelred}{rgb}{0.76, 0.23, 0.13}
\definecolor{ao(english)}{rgb}{0.0, 0.5, 0.0}

\definecolor{darkpastelred}{rgb}{0.76, 0.23, 0.13}
\definecolor{ao(english)}{rgb}{0.0, 0.5, 0.0}

\definecolor{yellow}{RGB}{255,255,153}
\definecolor{grey}{RGB}{224,224,224}

\newboolean{showcomments}
\setboolean{showcomments}{true}
\ifthenelse{\boolean{showcomments}}
 { \newcommand{\mynote}[2]{
      \fbox{\bfseries\sffamily\scriptsize#1}
        {\small$\blacktriangleright$\textsf{\emph{#2}}$\blacktriangleleft$}}}
        { \newcommand{\mynote}[2]{}}

\definecolor{DarkOrange}{rgb}{0.8,0.3,0.0}
\definecolor{DarkCyan}{rgb}{0.0, 0.55, 0.55}
\definecolor{DarkCyel}{rgb}{1.0, 0.49, 0.0}
\definecolor{yellow-green}{rgb}{0.6, 0.8, 0.2}

\newcolumntype{?}{!{\vrule width 1pt}}

\newcommand{\tool}{\texttt{MultiSEM}\xspace}

\newcommand{\find}[1]{
\begin{tcolorbox}[leftrule=0.2mm,toprule=0mm,bottomrule=0mm,left=0.0pt,right=0pt,top=0pt,bottom=0pt]%[tile,size=fbox,boxsep=2mm,boxrule=0pt,top=0pt,bottom=0pt,borderline={0.5mm}{0pt}{black!70!white},colback=black!5!white]
\em #1
\end{tcolorbox}
}

\newcommand{\equalcontrib}{\thanks{Equal contribution}}
\newcommand{\corrauthor}{\thanks{Corresponding author}}

\usepackage{natbib}  % DO NOT CHANGE THIS AND DO NOT ADD ANY OPTIONS TO IT
\usepackage{caption} % DO NOT CHANGE THIS AND DO NOT ADD ANY OPTIONS TO IT
\usepackage{algorithm}
\usepackage{algorithmic}

\usepackage{newfloat}
\usepackage{listings}

\DeclareCaptionStyle{ruled}{labelfont=normalfont,labelsep=colon,strut=off} % DO NOT CHANGE THIS
\floatstyle{ruled}
\newfloat{listing}{tb}{lst}{}
\floatname{listing}{Listing}
\title{Multilevel Semantic Embedding of Software Patches: A Fine-to-Coarse Grained Approach Towards Security Patch Detection}
\author{
    Xunzhu Tang\textsuperscript{\rm 1}\corrauthor,
    Zhenghan Chen\textsuperscript{\rm 2}\equalcontrib,
    Saad Ezzini\textsuperscript{\rm 3},
    Haoye Tian\textsuperscript{\rm 1},
    Yewei Song\textsuperscript{\rm 1},
    Jacques KLEIN\textsuperscript{\rm 1},
    Tegawendé F. Bissyandé\textsuperscript{\rm 1}
}
\affiliations{
    \textsuperscript{\rm 1}University of Luxembourg, Luxembourg\\
    \textsuperscript{\rm 2}Peking University, Beijing, China\\
    \textsuperscript{\rm 3}Lancaster University, Lancaster, UK\\

    xunzhu.tang@uni.lu,
    1979282882@pku.edu.cn,
    s.ezzini@lancaster.ac.uk,
    haoye.tian@uni.lu,
    yewei.song@uni.lu,
    jacques.klein@uni.lu,
    tegawende.bissyande@uni.lu
}
\usepackage{bibentry}
\begin{document}

\maketitle

\begin{abstract}
The growth of open-source software has increased the risk of hidden vulnerabilities that can affect downstream software applications. This concern is further exacerbated by software vendors' practice of silently releasing security patches without explicit warnings or common vulnerability and exposure (CVE) notifications. This lack of transparency leaves users unaware of potential security threats, giving attackers an opportunity to take advantage of these vulnerabilities. In the complex landscape of software patches, grasping the nuanced semantics of a patch is vital for ensuring secure software maintenance. To address this challenge, we introduce a multilevel Semantic Embedder for security patch detection, termed \tool{}. This model harnesses word-centric vectors at a fine-grained level, emphasizing the significance of individual words, while the coarse-grained layer adopts entire code lines for vector representation, capturing the essence and interrelation of added or removed lines. We further enrich this representation by assimilating patch descriptions to obtain a holistic semantic portrait. This combination of multi-layered embeddings offers a robust representation, balancing word complexity, understanding code-line insights, and patch descriptions. Evaluating \tool{} for detecting patch security, our results demonstrate its superiority, outperforming state-of-the-art models with promising margins: a 22.46\% improvement on PatchDB and a 9.21\%  on SPI-DB in terms of the F1 metric. 
\end{abstract}

\section{Introduction}
The rapid growth of the open source software (OSS) ecosystem has led to unparalleled advancements in computer software development. However, as with every silver lining, there's a cloud; the increasing reliance on OSS has been accompanied by a dramatic surge in the number of vulnerabilities. According to the 2021 OSSRA report~\cite{synopsys2023}, while 98\% of codebases are now composed of open source components, a large 84\% of these codebases contain at least one open-source vulnerability. More concerning is the fact that 60\% of them face high-risk vulnerability threats. Exploiting these chinks in the armor, attackers have launched "N-day" attacks against unpatched software systems. One glaring example of this is the remote command execution vulnerability (CVE-2021-22205) disclosed in April 2021~\cite{nvd2021}. Seven months after its release, more than 30,000 unpatched GitLab servers that weren't fixed were hacked, making them sources for DDoS attacks. This scenario underscores the significance of timely software patching—a tried and tested countermeasure against "N-day" attacks. However, in reality, the sheer volume of patches, which range from feature additions and performance bug resolutions to security vulnerability fixes, can be overwhelming. As a consequence, software updates are frequently deferred due to the intricate workflow of collating, testing, validating, and scheduling patches~\cite{10.1145/2858036.2858303}. Given this backdrop, there's an urgent need to enable users and administrators to differentiate security patches from the myriad of other updates. Yet, the task isn't as straightforward as it appears. Not all security patches make it to the NVD or are clearly marked in the changelog. Some software vendors, owing to the subjective nature of patch management, opt to release security patches on the down low~\cite{wang2020empirical}. Such "silent" security patches pose a conundrum for users and system administrators, leaving them in the dark about the true security implications and consequently, the urgency to apply them.

The academic and industry communities have explored multiple avenues to identify security patches. Traditional methods have revolved around machine learning (ML) models that primarily utilize syntax features~\cite{wang2020empirical,tian2012identifying}. A more contemporary approach involves deploying recurrent neural networks (RNNs) to treat the patch code as sequential data~\cite{wang2021patchrnn,zhou2021spi}. However, these methods fall short in two key aspects: they overlook program semantics and suffer from high false-positive rates. While ML-based solutions often miss out on capturing the intricate dependencies between code statements, RNN-based models, although inspired by natural language processing (NLP) techniques, neglect the unique attributes of programming languages. The end result? Unacceptably high false-positive rates, as evidenced by two RNN-based models that registered rates of 11.6\% and 33.2\%, respectively. Given that a mere 6-10\% of all patches are security-centric~\cite{wang2021patchdb}, the impetus to reduce these false positives becomes even more pronounced.

In this project, we aim to address these challenges. Through a fine-to-coarse grained approach, we present a novel method (\tool{}) that not only detects security patches with a high degree of accuracy but also reduces false positive rates. Drawing upon multilevel semantic embedding techniques and capitalizing on the content-rich information contained within software patches, our solution represents a quantum leap in the realm of security patch detection.

Our contributions are as follows:
\begin{itemize}
    \item \textbf{Multilevel Semantic Embedding}: We introduce a novel multilevel semantic embedding technique tailored for software patches. This method captures both high-level and granular details of the patches, ensuring a more nuanced and accurate representation.
    \item \textbf{Fine-to-Coarse Grained Approach}: Our approach not only focuses on individual lines of code but also looks at the broader structure and flow of the patch. The multi-scale perspective objective is to enhance the accuracy of security patch detection.
    \item \textbf{Experimental Results:} We conduct exhaustive evaluations of our proposed method against state-of-the-art solutions. Our results demonstrate superior performance, particularly in reducing false positive rates, which has been a longstanding challenge in the field.
\end{itemize}
\section{Approach}
\begin{figure*}[!ht]
    \centering
    \includegraphics[width=1\linewidth]{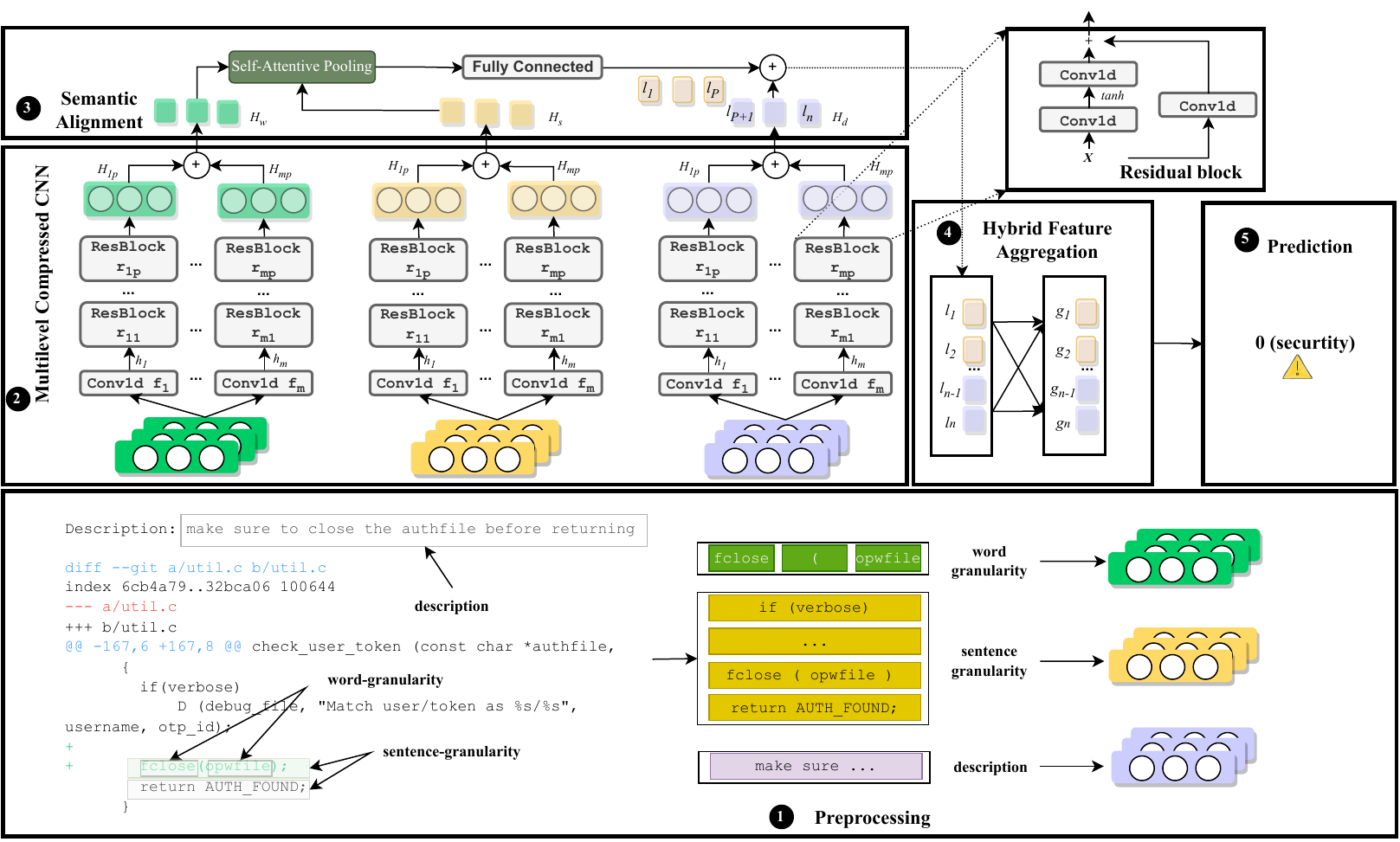}
    \caption{Architecture of \tool{}}
    \label{fig:architecture}
\end{figure*}

We present MultiSEM, a comprehensive framework comprising six distinct components: Preprocessing, Multilevel Compressed CNN, Semantic Alignment, Feature Refinement, Hybrid Feature Aggregation, and Prediction. Our model, designed to capture the fine-to-coarse-grained multilevel semantic embedding of software patches, takes the source code as its primary input and processes it through these components to yield a security patch detection output. The entire structure and flow of our approach can be visualized in Figure~\ref{fig:architecture}.

\subsection{Preprocessing}\label{sec:preprocessing}
As depicted in Figure~\ref{fig:architecture}(1), our first step is to decompose the input patches into three distinct segments: tokens, lines, and descriptions. At the word granularity, we define a word sequence \(w = \{w_1, w_2, \ldots, w_{nw}\}\), where \(nw\) represents the sequence length. We utilize the random embedding function from PyTorch to transform this sequence into a numerical vector. Hence, a word \(w_i\) translates to the vector \textbf{ew}$_i$. The cumulative word embedding of the patch is thus articulated as \(\textbf{EW} = \{\textbf{ew}_1, \textbf{ew}_2, \ldots, \textbf{ew}_{nw}\}\). Analogously, for sequence granularity, we represent sequence vectors as \(\textbf{ES} = \{\textbf{es}_1, \textbf{es}_2, \ldots, \textbf{es}_{ns}\}\), where \(ns\) indicates the number of lines, and the description vector is denoted by \(\textbf{ED} = \{\textbf{ed}_1, \textbf{ed}_2, \ldots, \textbf{ed}_{nd}\}\), with \(nd\) marking the sequence length of the description.

\subsection{Multilevel Compressed CNN (MCC)} \label{sec:multicompressed}

Within the domain of our multilevel approach, feature compression across all representational levels is pivotal. To this end, this section is bifurcated into two integral components: 
\begin{itemize}
    \item \textbf{Multi-Channel Convolutional Block}: Designed to harness contextual semantics inherent within various level representations.    
    \item \textbf{Compressed Residual Block}: A remedy to the vanishing or exploding gradient dilemma often encountered in deep networks, especially given the elongated nature of patches. The adoption of a residual structure enables gradients to directly traverse through residual connections, thereby substantially curtailing the associated risks.
\end{itemize}

\subsubsection{Multi Channel Convolutional Block}
To adeptly grasp patterns of varying lengths from our input, we adopt the multi-channel convolutional neural network as proposed in \cite{kim2014convolutional}. This approach utilizes filters, each characterized by distinct kernel sizes, which in essence, define the word window size. For an array of \(m\) channels, given as \(f_{1}, f_{2}, \ldots, f_{m}\), we correspondingly assign kernel sizes \(k_{1}, k_{2}, \ldots, k_{m}\). With these configurations, \(m\) 1-dimensional convolutions are executed on the input matrix \(E\). This convolutional operation can be mathematically described as:

\begin{equation}
    \begin{split}
    h_{i} &= \bigwedge_{j=1}^{n} \tanh \left(W_{i}^{T} E[j: j+k_{i}-1]\right),
    \end{split}
\end{equation}
where the symbol \(\bigwedge_{j=1}^{n}\) demarcates the convolutional operations performed in a word sequence. Crucially, the design choice ensures that the output word count \(n\) of \(h_{i}\) remains invariant with the input \(E\). This intention preserves the sequence length post-convolution. The term \(d^{f}\) signifies the out-channel size of a filter, with uniformity across filters in output dimensions. Delving deeper into the matrix details, \(E[j: j+k_{1}-1]\) =  and \(E[j: j+k_{m}-1]\) represent sub-matrices of \(E\). 

Therefore, after multi-channel convolutional block, we otain $h_1, \dots, h_m$ for word-level vectors, sentence-level vectors, and description-level vectors.

\subsubsection{Compressed Residual Block}

To refine the multi-channel convolved word embeddings, we incorporate a series of optimized residual blocks. These blocks offer not only a compact representation of features but also address potential challenges related to gradient dynamics, which is especially crucial for long word sequences.

\subsection{Residual Layer Overview}

The field of neural networks has witnessed significant advancements in recent years. One pivotal element that has consistently proven crucial in this evolution is the convolutional layer, responsible for primary feature extraction from input data. However, the challenge arises when we delve deeper: how can we maintain the hierarchical representation of features without the risk of information loss? An elegant solution, inspired by the work of He et al. \cite{he2016deep}, introduces the concept of Residual Blocks, visualized in Figure~\ref{fig:architecture}.

\paragraph{Architecture of a Residual Block}

For the residual block \( r_{m i} \), its architecture comprises three convolutional filters: \( r_{m i_{1}} \), \( r_{m i_{2}} \), and \( r_{m i_{3}} \). The computational procedure for these filters on the input can be articulated as:

\begin{equation}
    \begin{split}
    X_{1} &= r_{m i_{1}}(X) = \bigwedge_{j=1}^{n} \tanh \left(W_{m i_{1}}^{T} X[j: j+k_{m}-1]\right), \\
    X_{2} &= r_{m i_{2}}(X_{1}) = \bigwedge_{j=1}^{n} W_{m i_{2}}^{T} X_{1}[j: j+k_{m}-1], \\
    X_{3} &= r_{m i_{3}}(X) = \bigwedge_{j=1}^{n} W_{m i_{3}}^{T} X[j: j+k_{m}-1], \\
    H_{m i} &= \tanh \left(X_{2} + X_{3}\right),
    \end{split}
\end{equation}

In this context, \( X \) signifies the initial input to the block. Segments of this input, starting from the \( j \)-th row and concluding at the \( j+k_{m}-1 \)-th row, undergo transformations facilitated by the aforementioned filters.

% Upon closer examination of a typical residual block, denoted as \(r_{m i}\), it's evident that its architecture is predicated on three integral convolutional filters. These filters process the input in a sequential manner, where:

% \begin{equation}
% \begin{aligned}
% X_{1} &= \text{FirstFilter}(W_{m i_{1}}, X), \\
% X_{2} &= \text{SecondFilter}(W_{m i_{2}}, X_{1}),  \\
% X_{3} &= \text{ThirdFilter}(W_{m i_{3}}, X),  \\
% \text{Output}_{m i} &= \text{Activation} \left(X_{2}+X_{3}\right), 
% \end{aligned}
% \end{equation}

% Here, \(X\) serves as the primary input to the block. Segments of this input, ranging from the \(j\)-th row to the \(j+k_{m}-1\)-th row, undergo various transformations throughout the block.

\paragraph{Dimensionality and Spatial Relationships}
The matrix \(H_{m i}\) represents the output of each block and conforms to dimensions \(\mathbb{R}^{n \times d^{i}}\). The parameters \(d^{i-1}\) and \(d^{i}\) are crucial as they depict the input and output channel sizes respectively. Consequently, the in-channel dimension for the initial block is identified as \(d^{f}\), whereas the concluding block corresponds to \(d^{p}\).

The convolutional filters, discerned by the weight matrices \(W_{m i_{1}}, W_{m i_{2}},\) and \(W_{m i_{3}}\), exhibit differential properties in terms of kernel sizes. Notably, while the first two filters align in kernel size with their counterpart in the multi-filter convolutional layer, the third filter distinguishes itself with a singular kernel size.

Summarizing the architecture, the output matrix \(H_{m p}\) stands as a testament to the intricate relationship between the convolutional layer's \(m\)-th filter and its series of residual blocks. With a total of \(m\) filters, the ultimate output is a composite of individual outputs, mathematically represented as $H = H_{1p}\oplus H_{2p} \oplus \dots \oplus H_{mp})$.

Thus, after the residual block, we obtain $Hw$, $Hs$, $Hd$ for word-level, sequence-level, and description-level vectors, respectively.

\subsection{Semantic Alignment (SA)}\label{sec:semantic}
%In neural representations, effectively merging different data granularities often provides deep insights. Here, we combine the semantics from \( Hw \) (word-level) and \( Hs \) (sequence-level) vectors, aiming for a balanced understanding between token-level details and broader sequence contexts. Our approach employs two main strategies: \textit{Self-Attentive Pooling} to weigh component importance, and \textit{Feature Refinement Layer} to refine the pooled outputs into a compact, informative representation.
In the vast realm of neural representations, it's often the harmonious interplay between different granularities of data that yields the most insightful results. The extraction of both coarse and fine embeddings from patch code is no exception. In this section, we explore the strategic fusion of the semantic nuances encapsulated in \(Hw\) (word-level) and \(Hs\) (sequence-level) vectors. By aligning and juxtaposing these embeddings, we aim to achieve a holistic understanding, allowing the model to seamlessly traverse between detailed token-level insights and broader sequence contexts. To facilitate this synthesis, our methodology is bifurcated into two main strategies: \textit{Self-Attentive Pooling} and \textit{Feature Refinement Layer}. The former hones in on the weighted importance of various components, while the latter serves to amalgamate and further process the pooled outputs, ensuring that the final representation is both compact and informative.
\subsubsection{Self-Attentive Pooling}
In our pursuit of an effective semantic fusion, we first amalgamate the vectors \(Hw\) and \(Hd\) to formulate the composite vector \(Hwd\). Leveraging this combined vector, we introduce an attention-influenced soft-pooling technique to adeptly harmonize and integrate the underlying semantics of \(Hw\) and \(Hd\).

For an exemplar vector, represented as \(hwd_{j}\), and its contingent neighboring vectors \(\left\{hwd_{j+1}, \cdots, hwd_{j+g-1}\right\}\), our approach begins by deducing the localized attention scores. This is articulated by:

\begin{equation}
\alpha_{j}^{i}=\mathbf{Hwd}_{\alpha}^{i} \mathbf{x}_{j}^{i}+b
\end{equation}

Subsequent to this, the softmax function is employed to yield:
\begin{equation}
    \begin{split}
    \left[\beta_{j}^{i}, \cdots, \beta_{j+g-1}^{i}\right] &= \operatorname{softmax}\left(\left[\alpha_{j}^{i}, \cdots, \alpha_{j+g-1}^{i}\right]\right) \\
    \textit{s.t.}\quad \alpha_{j}^{i} &= \frac{(\boldsymbol{l}_{j}^i\boldsymbol{W}^Q)(\boldsymbol{l}_{j+g-1}^i\boldsymbol{W}^K)^\text{T}}{\sqrt{d}}
    \end{split}
\end{equation}
where $d$ is the dimension of the hidden state, and $\boldsymbol{W}^Q\in\mathbb{R}^{d\times d_q}$, $\boldsymbol{W}^K\in\mathbb{R}^{d\times d_k}$, $\boldsymbol{W}^V\in\mathbb{R}^{d\times d_v}$ are the learnable parameters matrices of the self-attention component. Here, we follow the previous works~\citep{vaswani2017attention,zugner2021language} and set $d_q=d_k=d_v=d$.

% \begin{equation}
% \left[\beta_{j}^{i}, \cdots, \beta_{j+g-1}^{i}\right]=\operatorname{softmax}\left(\left[\alpha_{j}^{i}, \cdots, \alpha_{j+g-1}^{i}\right]\right)
% \end{equation}

In this context, both \(\mathbf{Hwd_{\alpha}^{i}}\) and \(b\) are discerned as modifiable parameters. Post this determination, we engage in a soft-pooling mechanism on the \(g\) embeddings, which gives rise to the succinct representation:

\begin{equation}
\mathbf{o}_{p}^{i}=\sum_{q=j}^{j+g-1} \beta_{q}^{i} \mathbf{x}_{q}^{i}
\end{equation}

By adopting this structured approach, the entirety of \(nw+ns\) representations undergoes a metamorphosis, culminating in \(P=\left\lceil\frac{nw+ns}{g}\right\rceil\) refreshed representations, succinctly denoted as \(\left\{\mathbf{o}_{1}^{i}, \mathbf{o}_{2}^{i}, \cdots, \mathbf{o}_{P}^{i}\right\}\).

\subsection{Feature Refinement and Embedding Synthesis}
To extract high-level features from the transformed representations \(\left\{\mathbf{o}_{1}^{i}, \mathbf{o}_{2}^{i}, \cdots, \mathbf{o}_{P}^{i}\right\}\), they are passed through a dense neural layer, serving as a bridge to the final output.

For each \(\mathbf{o}_{k}^{i}\), where \(k \in [1, P]\), the transformation is:

\begin{equation}
    \begin{split}
    \mathbf{y}_{k}^{i} &= \mathbf{W}_{fc} \mathbf{o}_{k}^{i} + \mathbf{b}_{fc} \\
    \textit{s.t.}\quad l_{k} &= \operatorname{ReLU}(\mathbf{y}_{k}^{i})
    \end{split}
\end{equation}

Here, \(\mathbf{W}_{fc}\) is the weight matrix and \(\mathbf{b}_{fc}\) is the bias vector of the fully connected layer. This results in the output vectors \(l_1, l_2, \ldots, l_P\), which contain the distilled semantic information. The final embedding is a concatenation: \[l_1, \dots, l_n = (l_1, \dots, l_P)\oplus Hd.\]

% To further refine the representations and extract high-level features, we channel the transformed representations \(\left\{\mathbf{o}_{1}^{i}, \mathbf{o}_{2}^{i}, \cdots, \mathbf{o}_{P}^{i}\right\}\) through a dense neural layer. This fully connected layer serves as a bridge, connecting the rich, attention-pooled features to the final output representations.

% Formally, for each representation \(\mathbf{o}_{k}^{i}\), where \(k \in [1, P]\), we apply the following transformation:

% \begin{equation}
% \mathbf{y}_{k}^{i} = \mathbf{W}_{fc} \mathbf{o}_{k}^{i} + \mathbf{b}_{fc}
% \end{equation}
% where \(\mathbf{W}_{fc}\) represents the weight matrix and \(\mathbf{b}_{fc}\) denotes the bias vector of the fully connected layer. Subsequent to the transformation, an activation function (e.g., ReLU) is applied to \(\mathbf{y}_{k}^{i}\) to introduce non-linearity:

% \begin{equation}
% l_{k} = \operatorname{ReLU}(\mathbf{y}_{k}^{i})
% \end{equation}

% As a result of these operations, we are endowed with a series of output vectors: \(l_1, l_2, \ldots, l_P\), which encapsulate the distilled semantic information, ready for further tasks or interpretations.

% Finally, we concatenate \(l_1, l_2, \ldots, l_P\) and \(Hd\) to obtain the patch embedding with description semantics. The resulting concatenation is given by:
% \begin{equation}
% (l_1, l_2, \ldots, l_n) = \text{Concatenate}(l_1, l_2, \ldots, l_P, Hd).
% \end{equation}

\subsection{Hybrid Feature Aggregation through Advanced Attention Mechanism (HFA)}

In modern natural language processing and code understanding tasks, representing data with feature vectors plays a pivotal role. Among the concatenations derived from our model, the vector \((l_1, l_2, \ldots, l_n)\) stands out. This particular vector emerges from the fusion of two distinct embeddings, serving as a primary source of local feature representations. %The intrinsic nature of this concatenation, though enriched with data, tends to lean predominantly on its primary sources, thereby possibly sidelining the intricate integration between the code and description-level embeddings.

Given the nuanced interplay of these embeddings, a simple aggregation might not suffice.
%It is, therefore, imperative to devise a mechanism that takes into account both local subtleties and global contexts. 
Herein, the key-query attention mechanism presents itself as an optimal solution. Not only does it weigh the importance of each feature in the local context, but it also juxtaposes it against a broader, global context, resulting in a more balanced and informative feature representation.

Mathematically, using the attention mechanism, the global features are articulated as:
\begin{equation}
    \begin{split}
    \boldsymbol{g}_i = \sum\limits_{j=1}^{n}&\frac{\exp(\beta_{ij})}{\sum_{k=1}^n \exp(\beta_{ik})}(\boldsymbol{x}_j\boldsymbol{W}^V) \\
    \textit{s.t.}\quad \beta_{ij} &= \frac{(\boldsymbol{l}_i\boldsymbol{W}^Q)(\boldsymbol{l}_j\boldsymbol{W}^K)^\text{T}}{\sqrt{d}}
    \end{split}
\end{equation}
where $\boldsymbol{G}=[\boldsymbol{g}_1, \dots, \boldsymbol{g}_P]$. $\boldsymbol{x}_i\in\mathbb{R}^{d}$.

Through this approach, each compressed word window in our dataset not only retains its inherent local features but also gets enriched by the global context, thereby enhancing the overall representational power of our model.

\subsection{Binary Prediction Layer}

For the global vector \(\mathbf{D}_{g}\), a binary prediction is made with the sigmoid function:

\begin{equation}
    \begin{split}
    \tilde{y} &= \sigma(\mathbf{w}^{\top} \mathbf{D}_{g} + b) \\
    \textit{s.t.}\quad \tilde{y} &= \left(1 + \exp \left(-\mathbf{w}^{\top} \mathbf{D}_{g} - b\right)\right)^{-1},
    \end{split}
\end{equation}

Here, \(\mathbf{w}\) represents the learnable weight vector, and \(b\) denotes the bias term. The cross-entropy loss for binary prediction is:

\begin{equation}
    \begin{split}
    \mathcal{L} &= -y \log(\tilde{y}) - (1-y) \log(1-\tilde{y}),
    \end{split}
\end{equation}

Where \(y\) is the true label, and \(\tilde{y}\) is the predicted probability. This loss guides the optimization of the model's parameters.

\section{Experiment Design}
In this section, we present our experimental dataset, metrics, state-of-the-art, and research questions.
\subsection{Dataset}
We evaluate \tool{} on two popular datasets on security patch detection: PatchDB~\cite{wang2021patchdb} and SPI-DB~\cite{zhou2021spi}. 

\noindent
\textbf{PatchDB} offers a diverse collection of patches in C/C++ with 12K security-specific and 24K general patches. This dataset is an amalgamation of patches sourced from NVD reference links and direct GitHub commits from 311 renowned open-source projects, such as Linux kernel, MySQL, and OpenSSL. Such diversity allows us to comprehensively evaluate the robustness of security patch detection across various projects.

\noindent
\textbf{SPI-DB} focuses on patches from projects like Linux, FFmpeg, Wireshark, and QEMU. However, only data concerning FFmpeg and QEMU, totaling 25,790 patches (10K security and 15K non-security), has been made public. Together, these datasets not only grant us a wide spectrum of patch types for cross-project and within-project evaluation but also ensure a balanced representation, catering to both real-world applicability and optimal model training.

\subsection{Metric}
\noindent
{\bf +Recall} and {\bf -Recall} as presented in~\cite{tian2022change}, serve as specific metrics for evaluating patch correctness. The +Recall metric gauges the ability to predict correct patches, while -Recall assesses the effectiveness in filtering out incorrect ones.

\noindent
{\bf Area Under Curve (AUC) and F1 Score}. For the purpose of determining patch correctness, we developed a deep learning-based NLP classifier. To assess the effectiveness of our methodology, we employed the widely-recognized metrics: AUC and the F1 score. The latter is the harmonic mean of precision and recall, specifically applied for identifying correct patches~\cite{hossin2015review}. 

\noindent
{\bf True-Positive Rate (TPR)}. Also known as sensitivity, TPR quantifies how well a classifier identifies positive instances. In the context of security patches, it represents the percentage of valid patches correctly recognized. A high TPR indicates fewer overlooked genuine patches, reducing potential vulnerabilities~\cite{wang2023graphspd}.

\subsection{State-of-the-art}
\textbf{TwinRNN}: TwinRNN is metioned in \cite{wang2023graphspd}, emerging from the insights presented in \cite{zhou2021spi} and \cite{wang2021patchrnn}, the TwinRNN model leverages a unique architecture anchored on RNN-based solutions for detecting security patches. The model's moniker, "twin", originates from its dual RNN module setup, wherein each module processes pre-patch and post-patch code sequences, respectively. 

\noindent
\textbf{GraphSPD}: While TwinRNN offers a commendable benchmark in the realm of patch detection, GraphSPD, presents an alternative perspective and methodology. 

\subsection{Research Questions}
\textbf{RQ-1} How effective is {\em \tool{}} in security patch detection? \\
\textbf{RQ-2} How does {\em \tool{}} fare across various patch categories? \\
\textbf{RQ-3} What is the impact of key design choices on the performance of {\em \tool{}}?

\section{Experiment Results}
\subsection{[RQ-1:] Overall Performance}
\begin{table}[h!]
    \centering
    \caption{Comparison of TwinRNN, GraphSPD, and \tool{} on PatchDB and SPI-DB with various metrics (\%).}
    \resizebox{1\linewidth}{!}{
    \label{tab:comparison}
    \begin{tabular}{c|c|c|c|c|c|c}
        \toprule
        Method & Dataset & AUC & F1 & Recall+ & Recall-& TPR \\
        \midrule
        \multirow{2}{*}{$\begin{array}{c}\text{TwinRNN} \\
        \text{\cite{wang2021patchrnn}}\end{array}$} & PatchDB & 66.50 & 45.12 & 46.35 & 54.37 & 50.67\\
        & SPI-DB & 55.10 & 47.25 & 48.00 & 52.10 & 50.60 \\
        \hline 
        \multirow{2}{*}{$\begin{array}{c}\text{GraphSPD} \\
        \text{\cite{wang2023graphspd}}\end{array}$} & PatchDB & 78.29 & 54.73 & 75.17 & 79.67 & 70.82 \\
        & SPI-DB & 63.04 & 48.42 & 60.29 & 65.33 & 65.93 \\
        \hline 
        \multirow{2}{*}{\tool{}} & PatchDB & \cellcolor{black!25}83.15 & \cellcolor{black!25}77.19 & \cellcolor{black!25}79.52 & \cellcolor{black!25}86.78 &\cellcolor{black!25}79.52  \\
        & SPI-DB & \cellcolor{black!25}68.45 & \cellcolor{black!25}57.63 & \cellcolor{black!25}70.24 & \cellcolor{black!25}80.12 & \cellcolor{black!25}73.25 \\
        \bottomrule
    \end{tabular}}
\end{table}

\paragraph{Overall Results}
As depicted in Table \ref{tab:comparison}, our proposed method \textbf{\tool{}} yields the highest performance on both PatchDB and SPI-DB datasets. On PatchDB, \tool{} achieves an impressive AUC of $83.15\%$ with a commendable F1-score of $77.19$. For the SPI-DB dataset, \tool{} attains an AUC of $68.45\%$ with an F1-score of $57.63$. It's pertinent to note that the performances on PatchDB and SPI-DB shouldn't be directly compared due to their differing data distributions. Both datasets serve as our baseline to contrast our solution with existing methodologies.

\paragraph{Comparison with Security Patch Detection Approaches}
We assess the efficiency of \tool{} against both GraphSPD and TwinRNN by consistently applying the same training and test set divisions, as summarized in Table \ref{tab:comparison}.

\textbf{Effectiveness.} On the PatchDB dataset, \tool{} surpasses TwinRNN by a significant increase in AUC and an impressive increase in F1-score. When matched against GraphSPD, \tool{} remains superior in both AUC and F1-score. Similarly, for the SPI-DB dataset, our method \tool{} outperforms both TwinRNN and GraphSPD in AUC and F1-score. An enhanced AUC indicates that \tool{} has improved in distinguishing between positive and negative classes. The increased F1-score suggests that \tool{} provides a more balanced classification, successfully improving both precision and recall. The improvements in Recall+ and TPR indicate that our model is becoming more adept at correctly identifying positive instances, while the advancement in Recall- denotes better classification of negative instances.

\textbf{Practicality.} Precision and the false positive rate are pivotal metrics for ensuring reduced update frequencies and heightened labor productivity. As demonstrated in Table \ref{tab:comparison}, \tool{} reveals that a significant percentage of the predicted security patches are genuinely security-related. Moreover, for the SPI-DB dataset, \tool{} consistently excels in precision and has a reduced false positive rate, rendering it an optimal and pragmatic choice for real-world applications.

\find{{\bf \ding{45} Answer to RQ-1: }
Compared to previous approaches, the \textbf{\tool{}} method shows significant advancements: an improvement of 22.46\% in F1 over GraphSPD on PatchDB and 9.21\% on SPI-DB. This marked enhancement solidifies \textbf{\tool{}} as the optimal choice for real-world applications. }

\subsection{[RQ-2:] Assessing the Efficacy of {\em \tool{}} across Diverse Patch Categories}

\begin{table}[H]
    \centering
    \caption{Data statistics of PatchDB based on vulnerability type.}
    \label{tab:patchDB_stats}
    \resizebox{0.9\linewidth}{!}{
    \begin{tabular}{c|l|l|r}
        \hline
        \textbf{Severity} & \textbf{Vulnerability Type of Patch} & \textbf{Number} &\textbf{Proportion} \\
        \hline
        1 & Buffer overflow & 1211 &10.03\% \\
        2 & Improper authentication & 38 &0.31\% \\
        3 & Resource leakage & 197 &1.63\% \\
        4 & Double free/use after free & 1162 &9.62\% \\
        5 & Integer overflow & 602 &4.99\% \\
        6 & NULL pointer dereference & 8484 &70.27\% \\
        7 & Improper input validation & 155 &1.28\% \\
        8 & Uncontrolled resource consumption & 9 &0.07\% \\
        9 & Race condition & 25 &0.21\% \\
        10 & Uninitialized use & 62 &0.51\% \\
        11 & Other vulnerabilities & 128 &1.06\% \\
        \hline
    \end{tabular}}
\end{table}
PatchDB is a comprehensive dataset detailing various vulnerability types observed in software patches. We manually label all patches according to the types of resolved vulnerabilities. As shown in Table~\ref{tab:patchDB_stats}, the most prevalent vulnerability in the dataset is the "NULL pointer dereference," constituting a significant 70.27\% of the entries. Other notable vulnerabilities include "buffer overflow" at 10.03\% and "double free/use after free" making up 9.62\%. The dataset also captures more nuanced vulnerabilities, such as "improper authentication" and "uncontrolled resource consumption," representing 0.31\% and 0.07\%, respectively. Lesser observed vulnerabilities like "race condition" and "uninitialized use" are also cataloged, making PatchDB a diverse repository for analyzing and understanding software vulnerabilities. More details about labelling will be shown in Appendix.

\begin{figure}[H]
    \centering
    \includegraphics[width=\linewidth]{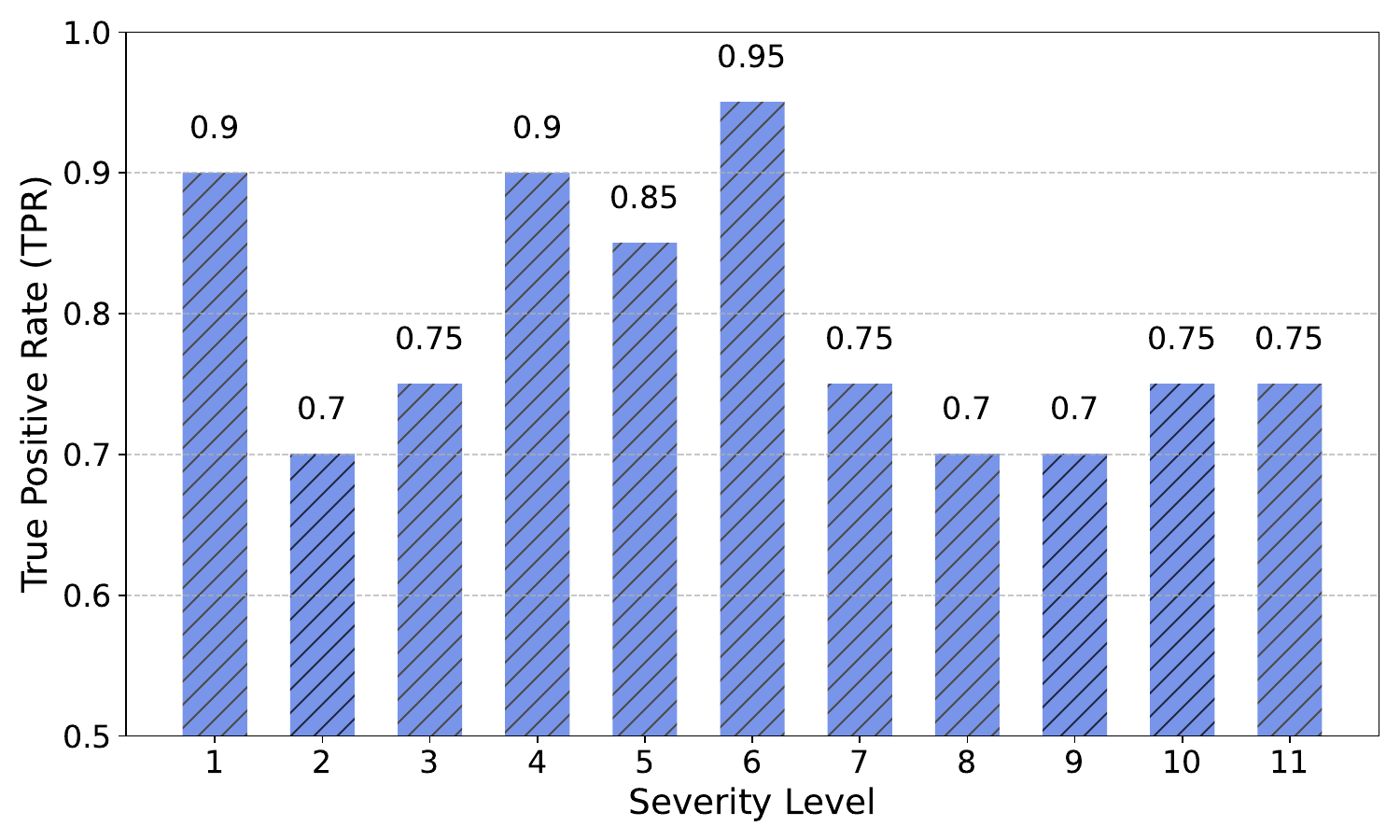}
    \caption{True Positive Rate (TPR) by Vulnerability Type in PatchDB Dataset.}
    \label{fig:tpr}
\end{figure}

As shown in Figure~\ref{fig:tpr}, through an in-depth performance analysis of \tool across various types of security patches, we unearthed two pivotal insights.

Firstly, security patch types with a commendable TPR by our tool, such as 'NULL pointer dereference' (with a TPR of 95\%), 'Buffer overflow', and 'Double free/use after free' (both around 90\%), indicate that these patches possess distinct features from their non-security counterparts. This distinction aids in designing more effective detection systems. For example, patches addressing 'Resource leakage' typically involve memory reinitialization and file operations, pointing towards memory API interactions. Similarly, patches for 'Race conditions' predominantly leverage lock/unlock operations to synchronize processes or threads, making them intimately linked with lock APIs. 

Secondly, some security patch types have a less impressive TPR under \tool, usually because they involve subtler security check modifications. Patches for issues like 'Improper input validation', 'Buffer overflow', and 'Improper authentication' often rely on conditional statements to delineate operational boundaries. While such checks are typical security patch patterns, they can easily be confounded with non-security patterns. Developers frequently employ conditional constructs to introduce new functionalities for specific scenarios. Hence, gleaning insights from the broader context is quintessential for accurate detection by \tool. Additionally, the effect of data imbalance is palpable. For instance, patches for 'Uncontrolled resource consumption' constitute only about 0.07\% of the dataset, furnishing sparse patterns for effective deep learning. We believe the performance for such categories would see a considerable enhancement with richer data availability.

\paragraph{Tail Problem}
As shown in Table~\ref{tab:tpr_comparison}, in order to evaluate the effectiveness in solving tail problem, we compare our \tool{} and GraphSPD on severity \#2,\#8,\#9, and \#10. 

\begin{table}[H]
    \centering
    \caption{Comparison of \tool and GraphSPD on TPR for specific severities.}
    \label{tab:tpr_comparison}
    \resizebox{1\linewidth}{!}{
    \begin{tabular}{c|c|c}
        \toprule
        \textbf{Severity} & \textbf{\tool TPR (\%)} & \textbf{GraphSPD TPR (\%)} \\
        \midrule
        2 (Improper authentication) & 70 & 68 \\
        8 (Uncontrolled resource consumption) & 70 & 65 \\
        9 (Race condition) & 70 & 67 \\
        10 (Uninitialized use) & 75 & 69 \\
        \bottomrule
    \end{tabular}}
\end{table}

The so-called "tail problem" in machine learning is a challenge that arises when certain classes or types within a dataset are under-represented, resulting in suboptimal performance for models trained on such data. Typically, these rare categories, or the 'tail' classes, may get overshadowed by the more dominant or frequently occurring classes during the model's learning process.

To gauge the prowess of our solution, \tool{}, in effectively addressing the tail problem, we carried out an in-depth comparison between \tool{} and its counterpart, GraphSPD. Our comparative study primarily focused on the specific severities of \#2, \#8, \#9, and \#10—categories that are notably challenging due to their sparse occurrence in the dataset.

The results of our study, as highlighted in Table~\ref{tab:tpr_comparison}, indicate a noticeable edge that \tool boasts over GraphSPD. For severity 2 (Improper authentication), our model registers a TPR of 70\%, surpassing GraphSPD's 68\%. Similar superiority is observed for severity 8 (Uncontrolled resource consumption) and severity 9 (Race condition), where \tool's TPRs are 70\% and 70\% respectively, in contrast to GraphSPD's 65\% and 67\%. The most prominent difference is discerned in severity 10 (Uninitialized use), where \tool leads with a 75\% TPR, a notable 6\% ahead of GraphSPD.

This comparative analysis underlines the adeptness of \tool in handling the tail problem. By achieving higher True Positive Rates (TPRs) across these challenging severities, \tool proves its efficiency in identifying and correctly classifying under-represented vulnerability types. This is crucial, as addressing the tail problem ensures that even the rarest of vulnerabilities do not slip under the radar, bolstering the overall security robustness.

\find{{\bf \ding{45} Answer to RQ-2: }
PatchDB highlights a variety of software vulnerabilities, with "NULL pointer dereference" being the most prevalent at 70.27\%. \textbf{\tool} performs exceptionally well in detecting major vulnerabilities but faces challenges with subtler ones due to data imbalances. In addressing the "tail problem" of under-represented classes, \textbf{\tool} consistently outperforms GraphSPD, exemplifying its capability to effectively detect even rare vulnerabilities. }

\subsection{[RQ-3:] Ablation Study}
% We investigate the related contribution of {\em token-level (\textbf{TL})}, {\em sentence-level (\textbf{SL})}, and {\em description-level (\textbf{DL})} by building three variants of \linebreak \tool where we remove either  {\em \textbf{TL}}  (\ie denoted as \tool$_{TL-}$), or  {\em \textbf{SL}} (\ie denoted as \linebreak \tool$_{SL-}$), or {\em \textbf{DL}} (\ie denoted as \linebreak \tool$_{DL-}$). We evaluate the performance of these variants on the task of security patch detection.

To discern the relative importance of the different levels of context used in our approach — specifically token-level (\textbf{TL}), sentence-level (\textbf{SL}), and description-level (\textbf{DL}) — we embarked on an ablation study. By systematically omitting one of these levels at a time, we generated three variants of \tool: \tool${TL-}$ (without token-level context), \tool${SL-}$ (sans sentence-level context), and \tool$_{DL-}$ (devoid of description-level context). The goal of this study was to shed light on how each contextual level contributes to the overall performance in security patch detection.

\begin{table}[H]
    \centering
    \caption{Performance evaluation of \tool variants on security patch detection (\%).}
    \label{tab:ablationstudy}
    \resizebox{1\linewidth}{!}{
    \begin{tabular}{c|c|c|c|c|c|c}
        \toprule
        Method & Dataset & AUC & F1 & Recall+ & Recall-& TPR \\
        \midrule
        \multirow{2}{*}{\tool$_{TL-}$} & PatchDB & 80.50 & 73.12 & 75.35 & 82.37 & 75.42\\
        & SPI-DB & 65.00 & 54.25 & 66.00 & 75.10 & 68.60 \\
        \hline 
        \multirow{2}{*}{\tool$_{SL-}$} & PatchDB & 81.50 & 74.00 & 76.20 & 83.40 & 76.50 \\
        & SPI-DB & 66.50 & 55.30 & 67.20 & 76.20 & 69.60 \\
        \hline 
        \multirow{2}{*}{\tool$_{DL-}$} & PatchDB & 78.00 & 70.50 & 72.00 & 80.00 & 73.00 \\
        & SPI-DB & 62.50 & 52.00 & 63.00 & 72.00 & 66.50 \\
        \hline 
        \multirow{2}{*}{\tool{}} & PatchDB & 83.15 & 77.19 & 79.52 & 86.78 &79.52  \\
        & SPI-DB & 68.45 & 57.63 & 70.24 & 80.12 & 73.25 \\
        \bottomrule
    \end{tabular}}
\end{table}

As revealed by the results in Table~\ref{tab:ablationstudy}: The removal of token-level information (\tool$_{TL-}$) significantly hampered the performance across both datasets, but it was especially pronounced in the PatchDB dataset. This indicates that token-level insights are vital, providing a granularity of detail that's essential for detecting nuances in security patches. Without the sentence-level context (\tool${SL-}$), there was a noticeable drop in performance, although not as drastic as with the \tool${TL-}$ variant. This reveals the utility of understanding the broader semantics of the code within the scope of a sentence or statement. This context helps in capturing relations between various tokens and offers a more comprehensive view than tokens in isolation. Most significantly, the exclusion of the description-level context (\tool$_{DL-}$) led to the most substantial degradation in performance. This was particularly evident in the AUC, F1, and Recall+ metrics across both datasets. Such a marked decline underscores the criticality of understanding the overarching narrative or intention behind a patch. The descriptive context often contains rich semantic information that can offer valuable hints or differentiate between patches, more so than local contexts like tokens or sentences.

In summation, while all contextual levels contribute positively to the performance, the description-level context emerged as the most influential. It proves that an understanding of the holistic description or rationale behind a patch is paramount in security patch detection tasks. This ablation study, thus, underscores the multi-faceted nature of our approach and reaffirms the necessity of a multi-level contextual understanding for high-precision security patch detection.

\find{{\bf \ding{45} Answer to RQ-3: }
The ablation study on \tool reveals the significant role each contextual level plays in security patch detection. The removal of the description-level context (\tool$_{DL-}$) resulted in the most pronounced performance drop, emphasizing its paramount importance in understanding patches. While all contexts are beneficial, the holistic understanding provided by the description-level is crucial for precise security patch detection. }
\section{Related Work}
\subsection{Advancements and Techniques in Security Patch Analysis}
In the realm of patch analysis, Li et al.~\cite{li2017large} undertook an empirical study of security patches, unearthing key behaviors. Soto et al.~\cite{soto2016deeper} provided insights into Java patches, advancing automated code repairs. VCCFinder~\cite{perl2015vccfinder} utilized SVM to detect suspicious patches, while Tian et al.~\cite{tian2012identifying} targeted bug corrections within Linux. SPIDER~\cite{machiry2020spider} highlighted secure patches that maintain normal program function. Rule-driven approaches for discerning security patches were proposed by Wu et al. and Huang et al.~\cite{wu2020precisely, huang2019using}. Vulmet~\cite{xu2020automatic} offers automatic urgent patching for Android, and Wang's group~\cite{wang2020machine} combined random forests with patch features to classify vulnerabilities. The rise in machine learning applications for patch analysis, especially deep learning, is evident in recent works~\cite{hoang2019patchnet,tian2020evaluating,hoang2019patchnet1}. PatchRNN~\cite{wang2021patchrnn} and SPI~\cite{zhou2021spi} employed RNNs for security patch identification, and GraphSPD~\cite{wang2023graphspd} tapped into graph structures to improve detection accuracy. The broader field of binary patch analysis includes binary differentiation~\cite{ming2017binsim,duan2020deepbindiff,zhao2020patchscope}, verification~\cite{dai2020bscout,zhang2021investigation}, recognition~\cite{xu2017spain}, and automation~\cite{duan2019automating,10.1145/3551349.3556914,niesler2021hera}.

\subsection{Progress in Sequential Data Methods}

The deep learning field has seen innovations in optimizing sequential data representations. Key strategies focus on multi-level methodologies~\cite{tang2021moto} capturing complexities in data from both NLP and CV. For example, Mototang et al.~\cite{mototang2021} applied sequence embeddings across sentence facets. Niu et al.~\cite{niu2020improving} introduced the MIA model for person re-identification. Du et al.~\cite{du2020fine} proposed a method for fine-grained visual classification. Other notable works include~\cite{jin2020multi,ling2023cross,li2021structext,zhou2021differentiable}. Additionally, compression techniques, like Luo et al.'s approach~\cite{luo2021fusion} with multi-filter CNNs and Resnet, have emerged for encoding complex sequences.

\section{Conclusion}
In the backdrop of an increasing reliance on open source software and the subsequent surge in vulnerabilities, there's an urgent need for accurate security patch detection. Our study introduces a groundbreaking method, employing a fine-to-coarse grained approach combined with multilevel semantic embedding techniques. This novel approach has proven to be highly effective, as evidenced by our experimental results, demonstrating significant improvements in accuracy and a notable reduction in false positives. As the software landscape continues to evolve, our methodology stands out, offering a beacon of hope for addressing the pressing challenges of security patch detection.

\section{Acknowledgments}
This work is supported by the NATURAL project, which has received funding from the European Research Council (ERC) under the European Union’s Horizon 2020 research and innovation programme (grant No. 949014).

\bibliography{aaai24}
\end{document}